\documentclass[pdflatex,sn-mathphys-ay]{sn-jnl}

\usepackage{graphicx}%
\usepackage{multirow}%
\usepackage{amsmath,amssymb,amsfonts}%
\usepackage{amsthm}%
\usepackage{mathrsfs}%
\usepackage[title]{appendix}%
\usepackage{xcolor}%
\usepackage{textcomp}%
\usepackage{manyfoot}%
\usepackage{booktabs}%
\usepackage{algorithm}%
\usepackage{algorithmicx}%
\usepackage{algpseudocode}%
\usepackage{listings}%
\usepackage{placeins}%

\newcommand{\trg}{\mathcal{T}}
\newcommand{\src}{\mathcal{S}}
\newcommand{\BBeta}{\boldsymbol{\beta}}

\newcommand{\Btheta}{\boldsymbol{\theta}}
\algrenewcommand\algorithmicrequire{\textbf{Input:}}
\algrenewcommand\algorithmicensure{\textbf{Output:}}
\DeclareMathAlphabet{\mathbbold}{U}{bbold}{m}{n}

\begin{document}
\title[Article Title]{A Principled Approach to Bayesian Transfer Learning}


\author*[1, 2]{\fnm{Adam} \sur{Bretherton}}\email{adam.bretherton@hdr.qut.edu.au}

\author[4]{\fnm{Joshua} \sur{J. Bon}}\email{joshuajbon@gmail.com}

\author[1, 2, 3]{\fnm{David} \sur{J. Warne}}\email{david.warne@qut.edu.au}

\author[1, 2]{\fnm{Kerrie} \sur{Mengersen}}\email{k.mengersen@qut.edu.au}

\author[1, 2, 3]{\fnm{Christopher} \sur{Drovandi}}\email{c.drovandi@qut.edu.au}

\affil*[1]{\orgdiv{School of Mathematical Sciences, Faculty of Science}, \orgname{Queensland University of Technology}, \orgaddress{\city{Brisbane}, \country{Australia}}}

\affil*[2]{\orgdiv{Centre for Data Science}, \orgname{Queensland University of Technology}, \orgaddress{\city{Brisbane}, \country{Australia}}}

\affil[3]{\orgdiv{Centre of Excellence for the Mathematical Analysis of Cellular Systems}, \orgname{Queensland University of Technology}, \orgaddress{\city{Brisbane}, \country{Australia}}}

\affil[4]{\orgdiv{School of Mathematical Sciences}, \orgname{Adelaide University}, \orgaddress{\city{Adelaide}, \country{Australia}}}

\abstract{Updating \textit{a priori} information given some observed data is the core tenet of Bayesian inference.  Bayesian transfer learning extends this idea by incorporating information from a related dataset to improve the inference on the observed target dataset which may have been collected under slightly different settings. The use of related information can be useful when the target dataset is scarce, for example.  There exist various Bayesian transfer learning methods that decide how to incorporate the related data in different ways.   Unfortunately, there is no principled approach for comparing Bayesian transfer methods in real data settings.  Additionally, some Bayesian transfer learning methods, such as the so-called power prior approaches, rely on conjugacy or costly specialised techniques.  In this paper, we find an effective approach to compare Bayesian transfer learning methods is to apply leave-one-out cross validation on the target dataset.  Further, we introduce a new framework, \textit{transfer sequential Monte Carlo}, that efficiently implements power prior methods in an automated fashion.  We demonstrate the performance of our proposed methods in two comprehensive simulation studies.}

\maketitle

\section{Introduction}\label{sec1}

Exploiting data from existing studies to improve the speed and quality of inference can be attractive when new data are expensive to obtain or the study is time-critical.  In epidemiology, for example, incorporating information from a previous epidemic can significantly improve the efficacy and speed of public health responses after the emergence of a new disease variant \citep{hao2021hotspots}.  Further, by incorporating related prior information into a small scale study, with relatively few data points, we can improve the quality of inference \citep{roster2022forecasting}.  Alternatively, compiling a small number of related studies could provide more robust inferences on the system of interest \citep{yao2010boosting}.

Under the Bayesian inference framework, we seek to improve inference based on a new dataset, referred to as the target, by incorporating information from related studies, referred to as the source, into the prior distribution.  Unfortunately, it is often not clear when and how to incorporate this source data in practice. Combining source and target data in the standard Bayesian way, often referred to as \textit{Bayesian updating}, may lead to inaccurate inference of model parameters when the true parameter values differ in the underlying the source and target datasets.  However, if the true parameter values from the source and target datasets are similar, it is inefficient to completely discard the source data.  Therefore, we might turn to so-called Bayesian transfer learning approaches to incorporate related information while avoiding (or reducing) any negative effects, such as bias, from the transferred information. Such effects are are difficult to identify in practice.  

There are several approaches to Bayesian transfer learning that are generally applicable to statistical models; the \textit{power prior} \citep{ibrahim2003optimality, ibrahim2012bayesian, ibrahim2015power}, the \textit{commensurate prior} \citep{hobbs2011hierarchical, hobbs2012commensurate, murray2014semiparametric} and the \textit{meta-analytic-predictive approach} \citep[MAPA,][]{neuenschwander2010summarizing, schmidli2014robust}.  Each of these methods incorporates the source data in a slightly different manner.

The commensurate prior incorporates the source data by allowing the parameters of interest to be perturbed versions of those in the source likelihood \citep{murray2014semiparametric}.  This perturbation allows the commensurate prior to model the bias between target and source parameters directly.  Unfortunately, the commensurate prior approach uses a spike and slab distribution in its setup which can be computationally prohibitive \citep{biswas2022scalable}.

Alternatively, the MAPA assumes the target and source model parameters are heterogeneous and treats them as different realisations from the same distribution.  Additionally, MAPA includes a robustness weight in its prior specification reducing how informative it is when the target and source data differ \citep{schmidli2014robust}.  However, this robust prior has a similar setup to the spike and slab prior and hence it is computationally costly to sample from the corresponding posterior.

In this work, we focus on the power prior and its variants.  The power prior is a class of informative priors that generalise Bayesian updating by tempering the likelihood of the source data with a transfer parameter $\alpha \in (0, 1)$.   When $\alpha=0$ the power prior recovers standard Bayesian inference on the target data and when $\alpha=1$ it is equivalent to Bayesian updating.  This transfer parameter can be treated as fixed, as in the \textit{fixed power prior} \citep[FPP,][]{ibrahim2015power}, or as random, as in the \textit{normalised power prior} \citep[NPP,][]{carvalho2021normalized, ye2022normalized}.  Selecting~$\alpha$ amounts to solving a model selection problem with a suitably chosen criterion.  In this paper, we use the \textit{model evidence} \citep{10.2307/2283136}. However, existing FPP approaches require re-computing the posterior for a grid of~$\alpha$ values.  The NPP creates a doubly intractable target distribution, since its normalising constant depends on $\alpha$ \citep{ye2022normalized, pawel2022normalized}.  One approach to overcome this is to use a conjugate prior on the model parameters \citep[e.g.][]{carvalho2021normalized}, but conjugate priors are only available for relatively simple statistical models.  \citet{park2018bayesian} avoid the need for conjugate priors by using specially designed Markov chain Monte Carlo (MCMC) algorithms for doubly intractable distributions, but these can be computationally intensive and require extensive tuning.



Despite the potential of Bayesian transfer learning, there are currently no principled approaches to compare the performance of different methods.    \citet{van2018including} and \citet{gravestock2019power} compare Bayesian transfer learning methods on simulated data where true parameter values are known, which are not available in real studies. For real case studies, they compare posterior summaries but do not provide a means to formally determine the best performing method.

In this work, we present three main contributions.  Firstly, we propose the use of posterior predictive checks on the target data to evaluate the performance of Bayesian transfer learning methods in real data settings where we do not have access to the true parameter values.  We find that \textit{leave-one-out cross-validation}  \citep[LOO-CV,][]{hastie2009elements} applied on the target data to be a suitable criterion for comparing methods since it reveals a similar ranking of the methods that is produced when the true parameter values are known. Secondly, we apply our evaluation framework to provide insight into the relative performance of various power prior methods under different simulation scenarios.  Although we only consider power prior methods in this paper, the criterion that we suggest can easily be applied to other Bayesian transfer learning methods.  Thirdly, we propose a new computational framework called \emph{transfer sequential Monte Carlo} (TSMC) that provides a computationally efficient and automated way for simultaneously implementing the FPP and NPP, and overcomes the intractable normalising constant issue of the latter approach.  We demonstrate the utility of our methods on two comprehensive simulation studies, a regression problem, and a survival analysis.


This paper proceeds as follows. In the next section, we provide further details on the variants of power priors.  In Section \ref{sec3} we introduce and rationalise the posterior predictive checks we use in our simulation studies. In Section \ref{sec4} we present the formulation of our TSMC framework.  The simulation studies and their results are described in Section \ref{sec5} and our findings are presented in Section \ref{sec6}.

\section{Bayesian Transfer Learning Methods}\label{sec2}
In this section, we provide a detailed overview of power prior methods.  We focus on power prior methods, as, in practice, we find that the commensurate prior and MAPA approaches are computationally prohibitive.  However, these two methods and the power prior similarly incorporate the source data to improve inference on the target.  Therefore, the posterior predictive checks on the target data we present in our simulation studies are applicable to all Bayesian transfer learning methods.

We first outline the notation used throughout this paper.  For simplicity, we consider the case with one source dataset, referred to as the source $\src$, which is related in some way to the dataset of interest, referred to as the target $\trg$.  Additionally, we assume the source and target likelihoods are from the same family of distributions. That is, all the target parameters $\Btheta_\trg$ in the space $\Theta$ have an equivalent source parameter $\Btheta_\src$ in the same space $\Theta$.  For clarity, $y$ denotes data with $y_\trg$ specifying the target data of size $n$ and $y_\src$ specifying source data of size $m$ with $n < m$.  The prior distribution for $\Btheta$ is denoted $\pi(\Btheta)$, the likelihood function for $\Btheta$ and $y$ is given by $p(y|\Btheta)$ and the associated posterior distribution is $\pi(\Btheta|y)$.  Below we discuss the different power prior approaches in detail.


\subsection{Power Prior}\label{subsec21}
Power prior methods are a generalization of Bayesian updating that reduces the influence of the source data by tempering the likelihood of the source data.  Specifically, the power prior uses the posterior of the source data tempered by $\alpha$ as the prior for the target model,
\begin{equation}\label{eq:sourcePP}
    \pi\left(\Btheta|y_\src, \alpha\right) = \frac{p(y_\src|\Btheta)^\alpha\pi(\Btheta)}{C_{\src}(\alpha)},
\end{equation}
where $C_{\src}(\alpha)$ is the normalising constant, which depends on the parameter $\alpha$.  Using Eq. \eqref{eq:sourcePP} as the prior for the target data results in the following posterior,
\begin{equation}\label{eq:targetPP}
    \pi\left(\Btheta| y_\trg, y_\src, \alpha\right) = \frac{p\left(y_\trg|\Btheta\right)\pi\left(\Btheta|y_\src, \alpha\right)}{C_{\trg}(\alpha)} = \frac{p\left(y_\trg|\Btheta\right)p(y_\src|\Btheta)^\alpha\pi(\Btheta)}{C_{\trg, \src}(\alpha)},
\end{equation}
where $C_{\trg}(\alpha)$ and $C_{\trg, \src}(\alpha)$ are normalising constants which depend on $\alpha$ and satisfy~$C_{\trg, \src}(\alpha) = C_{\trg}(\alpha)C_{\src}(\alpha)$.  In some areas of the power prior literature $C_{\src}(\alpha)$ is ignored \citep[e.g.][]{ibrahim2000power, chen2000power}.  However, \citet{ye2022normalized} show that without this term, Equation \eqref{eq:targetPP} violates the \textit{likelihood principle} \citep{birnbaum1962foundations} in $\alpha$.  Unfortunately, correctly including $C_\mathcal{S}(\alpha)$ results in a computationally challenging doubly intractable problem, due to the parameter dependent normalising constant. The common approach to deal with this challenge is through the use of conjugate priors \citep{gravestock2019power, carvalho2021normalized, ye2022normalized}.  In Section \ref{sec4} we take advantage of the identity $C_{\trg, \src}(\alpha) = C_{\trg}(\alpha)C_{\src}(\alpha)$ to target the correct Bayesian posterior in a computationally efficient manner.

Power prior methods are split into two categories. The first treats $\alpha$ as fixed and selects an optimal value according to  some model selection criterion \citep{ibrahim2003optimality, ibrahim2012bayesian, ibrahim2015power}.
There are several options for choosing a fixed value for $\alpha$ --- for example, the deviance information criterion \citep{ibrahim2012bayesian}. However, as the selection criterion must be calculated for every choice of $\alpha \in [0, 1]$ it can quickly become computationally intractable since a posterior must be computed for each value of $\alpha$.  We propose the use of model evidence as a selection criterion for $\alpha$ with our computationally efficient framework TSMC which we detail in Section \ref{sec4}.  The posterior for the FPP takes the following unnormalised form,
\begin{equation*}
    \pi_{\text{FPP}}\left(\Btheta | y_{\trg}, y_{\src}, \alpha\right) \propto p(y_{\trg} | \Btheta) \pi(\Btheta | y_{\src}, \alpha).
\end{equation*}
 Alternatively, $\alpha$ can be treated as random and assigned a prior \citep{duan2006evaluating, carvalho2021normalized, ye2022normalized} as is the case with the NPP.
The NPP takes a Bayesian approach to the power prior by assigning a prior to $\alpha \sim \text{ Beta}(\alpha_0, \beta_0)$ and correctly incorporates $C_{\src}$ with the following posterior,
\begin{equation}\label{eq:NPP}
    \pi_{\text{NPP}}\left(\Btheta, \alpha | y_\trg, y_\src\right) \propto p\left(y_\trg|\Btheta\right)\pi(\Btheta|y_\src, \alpha)\pi(\alpha) = p\left(y_\trg|\Btheta\right)\frac{p(y_\src|\Btheta)^\alpha\pi(\Btheta)}{C_{\src}(\alpha)}\pi(\alpha),
\end{equation}
where $\pi(\alpha)$ denotes the prior distribution for $\alpha$.  Unfortunately, it can be difficult to sample such posteriors in practice without using conjugate priors.  One can apply generic doubly intractable techniques for the NPP \citep[e.g.][]{park2018bayesian}, however, the FPP is not doubly intractable necessitating a separate method.  The efficient computational framework proposed in Section \ref{sec4} does not require a conjugate prior and can conveniently implement both the FPP and NPP.

\section{Comparing Bayesian Transfer Methods using Posterior Predictive Checks}\label{sec3}
Evaluating the accuracy of an estimated posterior under the Bayesian transfer learning setting can be difficult. Ideally, performance metrics would be based on a true parameter $\theta^*$.  Whilst this is suitable for simulation studies, $\theta^*$ is unavailable in real studies.  Here we develop performance metrics for comparing Bayesian transfer learning methods using carefully chosen posterior predictive checks.  In this section, we outline ideal metrics that we use to evaluate the performance of our proposed posterior predictive checks.

\subsection{Performance Metrics with Known $\theta^*$}

Ideally, we would use metrics such as the bias, posterior mean squared error (MSE) and a specified (say 90\%) frequentist coverage probability (FCP) of the true parameter value for each marginal parameter, $\theta$.  Together, these three metrics provide a detailed understanding of how accurately a posterior recovers the true parameter value $\theta^*$.  For this subsection we treat each component of $\theta$ separately.   

The bias is found by comparing the estimated posterior mean, $\hat{\mu}_{\theta} = \frac{1}{N}\sum_{i=1}^N\theta_i$, to~$\theta^*$,
\begin{equation*}\label{eq:bias}
    \text{Bias}\left(\{\theta_i\}_{i=1}^N, \theta^*\right) = \left|\hat{\mu}_{\theta} - \theta^*\right|,
\end{equation*}
where $\{\theta_i\}_{i=1}^N$ are the posterior samples obtained from a chosen Bayesian transfer learning method and $|\cdot|$ is the absolute value.  Similarly, the MSE is estimated by
\begin{equation*}\label{eq:mse}
    \text{MSE}\left(\{\theta_i\}_{i=1}^N, \theta^*\right) = \frac{\sum_{i=1}^N\left(\theta_i - \theta^*\right)^2}{N}.
\end{equation*}
The posterior samples are said to have $90\%$ FCP if $\theta^*$ is contained in $90\%$ of $90\%$ credible regions.  This requires multiple ($M$) datasets as for a single dataset the true value is either contained ($1$) or not contained ($0$) in the credible region.  We estimate the highest posterior density $90\%$ credible region for the $i$th posterior, $I^{(i)}_{0.9}$, using density estimation from the posterior samples \citep{hyndman1996computing}. Then the $90\%$ FCP is estimated by
\begin{equation*}\label{eq:cov}
    \text{FCP}_{0.9}\left(\{I^{(i)}_{0.9}\}_{i=1}^M, \theta^* \right) = \frac{1}{M}\sum_{i=1}^M\left(\mathbbold{1}\left\{\theta^* \in I^{(i)}_{0.9}\right\}\right),
\end{equation*}
where $\mathbbold{1}$ is an indicator function such that $\mathbbold{1}\{A\} = 1$ when $A$ is true  and $\mathbbold{1}\{A\} = 0$ otherwise.

\subsection{Performance Metrics with Unknown $\theta^*$}

The previous metrics require access to the true parameter value $\theta^*$.  Of course, in practice~$\theta^*$ is unknown, so other metrics must be devised.  In the Bayesian transfer learning setting we are interested in how well our method performs on the target data.  We propose the use of posterior predictive checks since they do not require $\theta^*$ and can be evaluated using only the target data.  Therefore, in our simulation studies in Section \ref{sec5} we evaluate the performance of two posterior predictive checks against these ideal metrics.

We first consider the \textit{expected log pointwise predictive density} \citep[ELPPD,][]{gelman2014understanding} given by,
\begin{equation*}
    \label{eq::elppd}
    \text{ELPPD} = \sum_{i=1}^{\tilde{N}} \mathbb{E}\left[\log p(\tilde{y}_i| y)\right],
\end{equation*} 
where $\tilde{y}$ is an out-of-sample dataset of size $\tilde{N}$ and $\pi(\tilde{y}|y)$ is the posterior predictive distribution.  In practice, we do not have access to an out-of-sample dataset and so must generate one.  We do this by separating the target data into a training set and a test set.  Unfortunately, under the Bayesian transfer learning setting, we often do not have access to enough data to form a separate test set.  As such, we must evaluate the posterior predictive of only the target data $y_\trg$.

To evaluate the expectation in the ELPPD on the target data we use a Monte Carlo approximation, which results in the \textit{computed log pointwise predictive density} \citep[CLPPD,][]{gelman2014understanding} estimated by,
\begin{equation*}
    \label{eq::clppd}
    \text{CLPPD}(\{y_{\trg, i}\}_{i=1}^n)  = \sum_{i=1}^n \log \left(\frac{1}{N}\sum_{j=1}^Np(y_{\trg, i}| \Btheta_j)\right),
\end{equation*}
where $\{\Btheta_j\}_{j=1}^N$ are the $N$ posterior samples from the chosen Bayesian transfer learning method.  However, we show empirically in Section \ref{sec5} that the performance of the CLPPD is poor, in that it does not align with the ideal metrics in terms of which transfer method performs the best. Since the CLPPD is evaluated on the target data, we find that this metric artificially boosts the performance of the transfer methods that most rely on the target data for fitting.    See Section \ref{sec5} for more discussion on why the CLPPD performs poorly in our Bayesian transfer learning context.


Therefore, we propose to use LOO-CV, so that each observation in the target dataset is tested, without being included in the dataset for inference and thus avoiding the overfitting problem that the CLPPD exhibits under the Bayesian transfer learning setting.  In place of a test set, LOO-CV instead repeatedly keeps a single data point as the current test and evaluates it with the posterior built from the remaining data, estimated by  
\begin{equation*}
    \text{LOO-CV}(\{y_{\trg, i}\}_{i=1}^n) = \sum_{i=1}^n \log \left(\frac{1}{N} \sum_{j=1}^N p\left(y_{\trg, i}|\Btheta_{(-i, j)}\right)\right),
\end{equation*}
where $\{\Btheta_{(-i, j)}\}_{j=1}^N$ are the $N$ posterior samples from the chosen Bayesian transfer learning method found without including $y_{\trg, i}$ in the target dataset.  The computational cost of building $n$ posteriors, even for small $n$, motivates the use of \textit{importance sampling} \citep{neal2001annealed,kahn1951estimation, kloek1978bayesian} to reduce this cost.  That is, for $y_{\trg, i}$ we reweight the $j$th sample from the Bayesian transfer posterior that includes all the target data as follows
\begin{equation*}
    \label{eq:NPPreweight}
    \begin{split}
        w_{-i}^{(j)} = \frac{p(y_{(\trg, -i)}|\Btheta_j)\pi(\Btheta_j|y_\src, \alpha)}{p(y_\trg|\Btheta_j)\pi(\Btheta_j|y_\src, \alpha)}
         = \frac{p(y_{(\trg, -i)}|\Btheta_j)}{p(y_\trg | \Btheta_j)},
    \end{split}
\end{equation*}
where $y_{(\trg, -i)}$ is the target data without the $i$th observation, which simplifies to~$w_{-i}^{(j)}~=~p(y_{(\trg, i)}|\Btheta_j)^{-1}$ if the observations are conditionally independent given $\Btheta$.  Then we take a weighted average inside our LOO-CV calculation to get the weighted LOO-CV (W-LOO-CV) as follows
\begin{equation*}
    \text{W-LOO-CV}(\{y_{\trg, i}\}_{i=1}^n) = \sum_{i=1}^n \log \left( \sum_{j=1}^N W_{-i}^{(j)}p\left(y_{\trg, i}|\Btheta_{(-i, j)}\right)\right),
\end{equation*}
where $W_{-i}^{(j)} = \frac{w_{-i}^{(j)}}{\sum_{k=1}^N w_{-i}^{(k)}}$. Alternative reweighting approaches could also be used, such as Pareto smoothed importance sampling \citep{vehtari2015pareto}, though these were not required in our examples and are not considered further here.

\FloatBarrier

\section{Transfer learning with Sequential Monte Carlo}\label{sec4}

In this section, we introduce our proposed \textit{transfer sequential Monte Carlo} (TSMC) framework for Bayesian transfer learning using the power prior. TSMC provides a computationally efficient approach to both the FPP and the NPP.  Additionally, the TSMC framework allows us to easily incorporate non-conjugate distributions with the power prior.  We achieve this by indirectly targeting the normalising constant $C_\trg(\alpha)$, employing the following decomposition
\begin{equation}\label{eq:decomp}
    \begin{split}
        C_{\trg}(\alpha) &= \frac{C_{\trg, \src}(\alpha)}{C_{\src}(\alpha)}.
    \end{split}
\end{equation}
We use $C_\trg(\alpha)$ to correctly normalise the conditional distribution shown in Eq. \eqref{eq:targetPP} and provide convenient access to the correct posterior for the FPP, found using model evidence, and the NPP, by assigning a prior to $\alpha$.  We use \textit{sequential Monte Carlo}  \citep[SMC,][]{chopin2002sequential, del2006sequential} extensively in our new framework.  Therefore, we introduce SMC algorithms in the next section before introducing TSMC.

\subsection{Sequential Monte Carlo}\label{subsec:smc} 
To efficiently estimate $C_{\trg, \src}(\alpha)$ and $C_{\src}(\alpha)$, we utilise an adaptive likelihood-annealing SMC algorithm \citep[e.g.][]{neal2001annealed, south2019sequential}.  SMC methods iteratively propagate a population of $N$ samples (particles) from an initial tractable distribution to a target distribution of interest.  Adaptive likelihood-annealing approaches connect the prior with the posterior by tempering the likelihood function $p(y| \Btheta)$ through a sequence of $t \in \{0, 1, \ldots, T\}$ distributions defined by an inverse temperature $\gamma_t$ such that $0 = \gamma_0 < \cdots < \gamma_T = 1$, with the $t$th distribution in the sequence
\begin{equation}\label{eq:smcdistro}
    \pi_t(\Btheta | y) \propto p(y| \Btheta)^{\gamma_t} \pi(\Btheta),
\end{equation} 
where $\pi(\Btheta)$ is the prior for $\Btheta$.

Assume we have a set of $N$ weighted particles from distribution $\pi_{t-1}(\Btheta|y)$, denoted as $\{W_{t-1}^{(i)}, \Btheta^{(i)}_{t-1}\}_{i=1}^N$, where $W_{t-1}^{(i)}$ is the normalised weight for the $i$th particle, $\Btheta^{(i)}_{t-1}$.  There are three main steps used to update from the distribution $t-1$ to the $t$th distribution.

First, the reweight step; this step reweights each particle according to the next distribution in the sequence using importance sampling. The unnormalised weight adjustment for the $i$th particle from the $t$th distribution is given by
\begin{equation*}
    w_{t}^{(i)} = W_{t-1}^{(i)}\cdot p(y|\Btheta^{(i)}_{t-1})^{(\gamma_t - \gamma_{t-1})},
\end{equation*}
which can then be appropriately normalised via
\begin{equation*}
    W_{t}^{(i)} = \frac{w_{t}^{(i)}}{\sum_{j=1}^N w_{t}^{(j)}}.
\end{equation*}
For a given target, with index $t$ omitted for notational simplicity, we can use the weights to compute the \textit{effective sample size} \citep[ESS,][]{del2006sequential} with
\begin{equation*}
    \text{ESS}\left(\{\Btheta^{(i)}, W^{(i)}\}_{i=1}^N\right) = \frac{1}{\sum_{i=1}^N (W^{(i)})^2}.
\end{equation*}
The next inverse temperature $\gamma_{t+1}$ is adaptively chosen such that the ESS for the $N$ particles from the $t$th distribution is approximately equal to some threshold $E$  (often set to $N/2$, which we do in this paper).

Secondly, the resample step updates the particle system to favour particles with high importance weights without altering the $t$th distribution.  Resampling duplicates particles with large weights and replaces particles with low weights, to better explore the high probability regions of the $t$th distribution.  Several resampling algorithms can be used, the most basic of which is multinomial resampling.  We use stratified resampling as it typically performs better than multinomial resampling \citep{kitagawa1996monte, gerber2019negative}.  Without resampling, the importance weights for the population of particles may concentrate on only a few particles.  After resampling occurs, the new population of particles will have uniform weights and may have duplicated particles.

Finally, the rejuvenate step; this step aims to perturb the particles using a small number of MCMC steps \citep[e.g. random walk Metropolis-Hastings with covariance matrix adapted using the current population of particles, see][]{chopin2002sequential} to mitigate degeneracy from particle duplicates.  The MCMC kernel is $\pi_t$-invariant, hence this perturbation rejuvenates the population of particles without altering the distribution they approximate.  Typically, a multivariate normal distribution is used for the proposal distribution 
\begin{equation*}
    q(\Btheta^*|\Btheta^{(i)}) = \mathcal{MN}(\Btheta^*; \Btheta^{(i)}, \Sigma),
\end{equation*}
where $\Sigma$ is a covariance matrix \citep{chopin2002sequential, jasra2011inference, south2019sequential}.  A convenient heuristic for $\Sigma$ is to use the sample covariance matrix computed from the current population of particles.

Often, a single MCMC step will not move every duplicate particle, necessitating multiple MCMC steps per iteration of the rejuvenation step.  There are two options for choosing the number of MCMC steps for the $t$th distribution $R_t$; either set $R_t$ to a fixed value or choose it adaptively.  The required number of MCMC steps to sufficiently diversify the population of particles may change for each of the $t$ distributions in the sequence.  Therefore, an adaptively chosen $R_t$ may be preferred.  \citet{south2019sequential} \citep[see also][]{drovandi2011estimation} show that $R_t$ can be chosen adaptively such that there is a $1-c$ probability that each particle has moved at least once with
\begin{equation*}
    R_t = \left\lceil \frac{\log(c)}{\log(1-\hat{p}_t)}\right\rceil,
\end{equation*}
where $\lceil\cdot\rceil$ is the ceiling function, and $\hat{p}_t$ is the acceptance rate of $S$ trial MCMC moves.  The reweight, resample and rejuvenate steps are repeated until the population of particles approximates the terminal distribution, Eq. \eqref{eq:smcdistro} when $\gamma=1$.

Using an SMC sampler, one can obtain a convenient estimator of the normalising constant of the target distribution \citep{del2006sequential}.  That is, we set the initial normalising constant $Z_{0}=1$, assuming initial particles drawn from a normalised prior, and use the estimated normalising constant update for the $t$th distribution given by
\begin{equation*}
    \frac{Z_t}{Z_{t-1}} \approx \sum_{i=1}^N  w_{t}^{(i)}.
\end{equation*}
Then we can estimate the normalising constant for the $t$th distribution via the identity,
\begin{equation}\label{eq:NormConstantEstim}
    Z_t = \frac{Z_t}{Z_0} = \frac{Z_t}{Z_{t-1}}\frac{Z_{t-1}}{Z_{t-2}} \cdots \frac{Z_1}{Z_0} = \prod_{k=1}^t \frac{Z_k}{Z_{k-1}}.
\end{equation}
Further, $Z_t$ is referred to as the model evidence when it is used as a selection criterion for the $t$th posterior distribution.

\subsection{Transfer Sequential Monte Carlo}\label{subsec:tsmc}
The conditional distribution in Eq. \eqref{eq:targetPP} is a key component of both the FPP and NPP and requires $C_\trg(\alpha)$ to be correctly normalised.  Estimating $C_\trg(\alpha) \ \forall \alpha \in [0, 1]$ directly is computationally prohibitive.  Therefore, we use the decomposition in Eq. \eqref{eq:decomp} to indirectly estimate $C_\trg(\alpha)$ by instead estimating $C_\src(\alpha)$ and $C_{\trg,\src}(\alpha)$.

First, we consider the SMC algorithm that we employ to estimate $C_\src(\alpha)$, schedule~$1$, which targets the $t$th posterior distribution given by
\begin{equation}\label{eq:CSpost}
    \pi_{t, \src}\left(\Btheta|y_\src, \alpha_t\right) \propto p(y_\src|\Btheta)^{\alpha_t}\pi(\Btheta),
\end{equation}
where $\alpha \in [0, 1]$ is the inverse temperature.  We adaptively select the inverse temperature schedule for the sequence of $t \in \{0, 1, \ldots, T\}$ distributions such that
\begin{equation*}
    0 = \alpha_0 < \alpha_1 < \cdots < \alpha_{T-1} < \alpha_T = 1,
\end{equation*}
ensuring that the ESS for the population of particles at the next inverse temperature is approximately equal to $E$.  The unnormalised weight adjustment for the $i$th particle in the $t$th posterior distribution is
\begin{equation*}
    w_{t}^{(i)} = W_{t-1}^{(i)}\cdot p(y_\src|\Btheta^{(i)})^{\alpha_t - \alpha_{t-1}},
\end{equation*}
where $W_{(0, i)} = 1/N$ for all $N$ particles.  We estimate $C_\src(\alpha_t)$ for each $\alpha_t$ in the sequence as in Eq. \eqref{eq:NormConstantEstim}.  To ensure that we can conveniently estimate $C_\src(\alpha) \ \forall \alpha \in [0, 1]$, we store the estimated $C_\src(\alpha_t)$ and the population of particles for each of the $T$ distributions described by Eq. \eqref{eq:CSpost}.

Next, we consider the estimate of $C_{\trg, \src}(\alpha)$ and the SMC algorithm we employ to estimate it, schedule $2$.  Schedule $2$ targets the $t$th posterior distribution given by
\begin{equation}\label{eq:smcpost}
    \pi_{t, \text{TSMC}}\left(\Btheta|y_\src, y_\trg, \gamma_t, \alpha_t\right) \propto p(y_\trg|\Btheta)^{\gamma_t}p(y_\src|\Btheta)^{\alpha_t}\pi(\Btheta),
\end{equation}
where $\gamma \in [0, 1]$ and $\alpha \in [0, 1]$ are the inverse temperatures.  With a carefully constructed sequence of $T^*$ distributions, we can efficiently incorporate both the target data and the source data such that we have the correct approximation of $C_{\trg, \src}(\alpha)$ for Eq. \eqref{eq:decomp}.  First, we use $\gamma$ to traverse from the prior to the target posterior using only the target data for $t \in \{0, 1, \ldots, T\}$.  Then we incorporate the source data for $t \in \{T, T+1, \ldots, T^*\}$ with $\alpha$.  That is, for $t\in \{0, 1, \ldots, T, T+1, \ldots, T^*\}$ we define the sequence of distributions by adaptively selecting the inverse temperatures
\begin{equation*}
    \begin{split}
        0 &= \gamma_0 < \gamma_1 < \cdots < \gamma_{T-1} < \gamma_T = \gamma_{T+1} = \cdots = \gamma_{T^*} = 1 \\
        0 &= \alpha_0 = \alpha_1 = \cdots = \alpha_{T-1} = \alpha_T < \alpha_{T+1} < \cdots < \alpha_{T^*} = 1,
    \end{split}
\end{equation*}
ensuring that the ESS for the population of particles at the next inverse temperature is approximately $E$.  The unnormalised weight adjustment for the $i$th particle in the $t$th distribution for schedule $2$ is given by,
\begin{equation*}
    w_{t}^{(i)} =
    \begin{cases}
        W_{t-1}^{(i)} \cdot p(y_{\trg}|\Btheta^{(i)})^{\gamma_{t} - \gamma_{t-1}} & \text{for } t \leq T\\
        W_{t-1}^{(i)} \cdot p(y_{\src}|\Btheta^{(i)})^{\alpha_{t} - \alpha_{t-1}} & \text{for } t > T,
    \end{cases}
\end{equation*}
where $W_{(0, i)} = 1/N$ for all $N$ particles.  As in schedule $1$, we estimate $C_{\trg, \src}(\alpha_t)$ at each $\alpha_t$ in the inverse temperature schedule and we store both the estimate of $C_{\trg, \src}(\alpha_t)$ and the population of particles for each of the $T$ distributions described by \eqref{eq:smcpost}, so that we can conveniently estimate $C_{\trg, \src}(\alpha_t) \ \forall \alpha \in [0, 1]$.

\begin{figure}[h]
    \centering{
    \resizebox{\textwidth}{!}{
        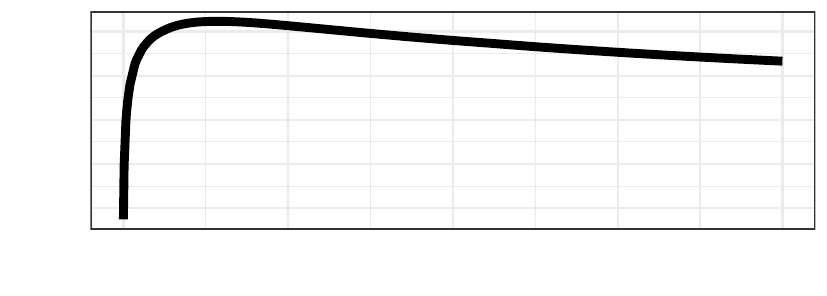
        }
    }
    \caption{Line plot between estimates of $\log C_\trg(\alpha)$ for $\alpha \in [0, 1]$ using a single run of TSMC on a dataset with moderate difference between source data and target data.}
    \label{fig:Cplot} 
\end{figure}

Our construction of two schedules and the decomposition in Eq. \eqref{eq:decomp} allows convenient access to any intermediate $\alpha \in (\alpha_t, \alpha_{t+1})$ through the use of importance sampling with a guaranteed ESS greater than $E$ and therefore substantially reduces the computational complexity of finding $C_\trg(\alpha) \ \forall \alpha \in [0, 1]$.  We use $N=1000$ particles for our experiments and find that the Monte Carlo variability is small enough that we can find the optimal estimates of $C_\trg(\alpha)$ with reasonable accuracy as shown in Figure \ref{fig:Cplot}.  Further, storing the population of particles from schedule $2$ provides convenient access to the conditional distribution shown in Eq. \eqref{eq:targetPP}, which both the FPP and the NPP require.

We propose a model selection approach for the FPP, \textit{transfer sequential Monte Carlo - model evidence} (TSMC-ME), which uses the SMC estimate of the model evidence $C_\trg(\alpha)$ to choose $\alpha$, with
\begin{equation}\label{eq:FPPcrit}
    \alpha^* = \underset{\alpha \in [0, 1]}{\arg\max} \ C_\trg(\alpha),
\end{equation}
which can be considered an empirical Bayes procedure \citep{maritz2018empirical}.  To ensure we choose the appropriate value for $\alpha$, we consider values outside of the inverse temperature schedule with a grid search algorithm.  This grid search algorithm chooses a set $G$ of equally spaced grid points on $(0, 1)$ and then approximates $C_\trg(\alpha) \ \forall \alpha \in G$ using importance sampling as necessary.  If the chosen $\alpha^*$ is in the original inverse temperature schedule, the posterior samples for the FPP are the stored samples from the associated conditional distribution shown in Eq. \eqref{eq:targetPP} found with schedule $2$.  However, if $\alpha^*$ is not in the original inverse temperature schedule, the posterior samples for the FPP are found using a single reweight, resample and rejuvenate step on the stored samples associated with the largest $\alpha < \alpha^*$ in the original inverse temperature schedule.  

To facilitate the fully Bayesian approach for the power prior, \textit{transfer sequential Monte Carlo - normalised power prior} (TSMC-NPP), we utilise the conveniently estimated values of $C_\trg(\alpha)$ to appropriately weight draws from the prior $\pi(\alpha)$ of $\alpha$ with


\begin{align}\label{eq:nppdecomp}
    \pi(\alpha|y_{\trg}, y_{\src}) &= \int_\Theta \pi(\Btheta, \alpha | y_{\trg}, y_{\src})d\Btheta \nonumber\\
     &\propto \int_\Theta \frac{p(y_{\trg}|\Btheta)p(y_{\src}|\Btheta)^\alpha \pi(\Btheta)}{C_{\src}(\alpha)}\pi(\alpha) d\Btheta \nonumber\\
     &\propto \frac{C_{\trg, \src}(\alpha)}{C_{\src}(\alpha)}\pi(\alpha) = C_{\trg}(\alpha)\pi(\alpha).
\end{align}
To draw a joint sample from the posterior in Eq. \eqref{eq:NPP}, we first draw $N$ samples from the prior $\{\alpha^{(i)}\}_{i=1}^N \sim \pi(\alpha)$.  Then we use the estimates of $C_\trg(\alpha^{(i)})$ as unnormalised weights, which we normalise to appropriately weight all $N$ samples, as in Eq. \eqref{eq:nppdecomp}.  Finally, for each $\alpha^{(i)}$ we draw a $\Btheta^{(i)}$ from the conditional distribution shown in Eq. \eqref{eq:targetPP}, noting that we have convenient access to estimates of both $C_{\trg}(\alpha)$ and Eq. \eqref{eq:targetPP} for any $\alpha \in [0, 1]$ with importance sampling and the stored posterior particles from both schedules.

The conditional distribution shown in Eq. \eqref{eq:targetPP} is used by both TSMC-ME and TSMC-NPP. Therefore, we recommend that each estimate of the conditional distribution be stored and that both posteriors be estimated and then compared with LOO-CV.  Additionally, TSMC-NPP should be performed first as the $N$ approximations of $C_\trg(\alpha)$ can be included in the set $G$ for TSMC-ME.  In Algorithm \ref{alg:tsmcPseudo}, we provide a meta-algorithm for the TSMC framework that returns a set of posterior samples for the FPP, using TSMC-ME, and the NPP, using TSMC-NPP. 

\begin{algorithm}
\caption{Transfer sequential Monte Carlo}
\label{alg:tsmcPseudo}
\begin{algorithmic}[1]
\Require The target data $y_\trg$, the source data $y_\src$, the target likelihood $p(y_\trg | \Btheta)$, the source likelihood $p(y_\src|\Btheta)$, the prior for $\Btheta$ $\pi(\Btheta)$, the prior for $\alpha$ $\pi(\alpha)$, the number of particles $N$ and the set of grid points $G$.
\Ensure The posterior particles for the NPP $\{(\Btheta^{(i)}, \alpha^{(i)})\}_{i=1}^N$ and the posterior particles with chosen $\alpha^*$ for the FPP $(\{\Btheta^{*(i)}\}_{i=1}^N, \alpha^*)$.
\State Approximate $\pi(\Btheta|y_\trg) \propto p(y_\trg|\Btheta)\pi(\Btheta)$ using SMC as in Section \ref{subsec:smc}.
\State Approximate $C_{\trg, \src}(\alpha), C_\src(\alpha)$ and Eq. \eqref{eq:targetPP} using two SMC schedules as in Section \ref{subsec:tsmc}.
\State Draw $N$ prior particles $\alpha^{(i)} \sim \pi(\alpha)$, estimate $C_\trg(\alpha^{(i)})$ for each and weight each $\alpha^{(i)}$.
\State Draw $\Btheta^{(i)}$ from conditional distribution in Eq. \eqref{eq:targetPP} for each $\alpha^{(i)}$.
\State Estimate $C_\trg(\alpha_G)$ for each $\alpha_G \in G$ using importance sampling and Eq. \eqref{eq:NormConstantEstim}.
\State Choose $\alpha^*$ with Eq. \eqref{eq:FPPcrit}
\If{$\alpha^*$ is in inverse temperature schedule}
    \State Retrieve stored $\{\Btheta^{*(i)}\}_{i=1}^N$ for associated $\alpha^*$. 
\Else
    \State Find $\{\Btheta^{*(i)}\}_{i=1}^N$ with a single reweight, resample and rejuvenate step as in Section \ref{subsec:smc}.
\EndIf
\RenewDocumentCommand{\alglinenumber}{ m }{} 
\State \Return $\{(\Btheta^{(i)}, \alpha^{(i)})\}_{i=1}^N$ for the NPP and $(\{\Btheta^{*(i)}\}_{i=1}^N, \alpha^*)$ for the FPP.
\end{algorithmic}
\end{algorithm}

\FloatBarrier

\section{Simulation Studies}\label{sec5}
We evaluate the performance of Bayesian transfer learning approaches using simulation studies that include two example models. The first example is a linear regression model, and the second is a Weibull cure model based on modelling used in melanoma cancer clinical trials \citep{kirkwood1996interferon, kirkwood2000high}. 

To simulate a Bayesian transfer setting, we generate three types of datasets for each simulation study using a set of shared covariates of size $n + m$, with $n < m$.  The first is the true dataset of size $n + m$, which we draw using the true parameter value and represents a dataset where there is no misspecification between the target and source.  The second is the target dataset of size $n$, which is a subset of the true dataset taking the first $n$ data points.  The third is the source dataset of size $m$, which we draw using a parameter value that is shifted from the true parameter value and the last $m$ covariate values.  We shift the true parameter value $2k$ standard deviations (according to the posterior found using the true dataset) across four levels, $k \in \{0, 1, 2, 3\}$, to represent increasing misspecification in the source dataset.

In these simulation studies, we evaluate six methods.  The first method is Bayesian inference on the target data only (BT).  The second method is Bayesian inference on the source data only (BS).  The third method is a standard Bayesian updating approach (BU).  The fourth method is a fixed power prior approach implemented using the TSMC framework with the model evidence as the selection criterion for $\alpha$.  The fifth method is a normalised power prior approach implemented using the TSMC framework with a Beta$(1, 1)$ prior for $\alpha$.  The sixth method is standard Bayesian inference on the true dataset, which reuses the covariates from the BT and BS methods (True). The True method is the gold standard, but generally unavailable, approach that we implement to facilitate comparisons with the other methods.  We note that when $k=0$, the BU and True methods are equivalent.

To determine the posterior accuracy of the competing methods and the validity of posterior predictive checks for identifying appropriate transfer, we report on the average of five comparison metrics over $100$ independently generated datasets for each particular transfer problem.  The five comparison metrics are the bias, posterior MSE, FCP, CLPPD and LOO-CV. We use the metrics to highlight the best performing transfer method, with reference to the posterior obtained from fitting to the $(n + m)$-sized dataset generated using the true parameter value (true posterior), as we increase the difference between source and target across four levels of $k$.  Additionally, since we find variability within CLPPD and LOO-CV values across the $100$ replicate experiments, we also consider ranking the transfer methods based on the respective CLPPD or LOO-CV values for a given dataset and then compute the average rank across the 100 replicate experiments.  A smaller average rank indicates better performance of the corresponding Bayesian transfer learning method.  The simulation study code is available at \url{https://github.com/Lemiltock/TSMC-SimStudy}.

\subsection{Example 1: Linear regression model}\label{subsec51}
For the first example, we consider a linear model with three model parameters $\beta_0, \beta_1,$ and $\sigma$, where a single observation $y$ is generated via
\begin{equation*}
    \begin{split}
        y &= \beta_0 + x \beta_1 + \epsilon \\
        \epsilon &\sim \mathcal{N}(0, \sigma^2),
    \end{split} 
\end{equation*}
where $y$ is the response, $\epsilon$ is the observation error which follows a normal distribution with standard deviation $\sigma$ and $x$ is the covariate value drawn from a $\mathcal{N}(0, 1)$ distribution.  To simulate two related datasets we draw a target dataset $\left\{y_{\trg, i}, x_{\trg, i}\right\}_{i=1}^n$ and source dataset $\left\{y_{\src, i}, x_{\src, i}\right\}_{i=1}^m$ of size $n=40$ and~$m=80$ with differing parameter values as follows.

\begin{figure}[h]
    \centering{
    \resizebox{\textwidth}{!}{
        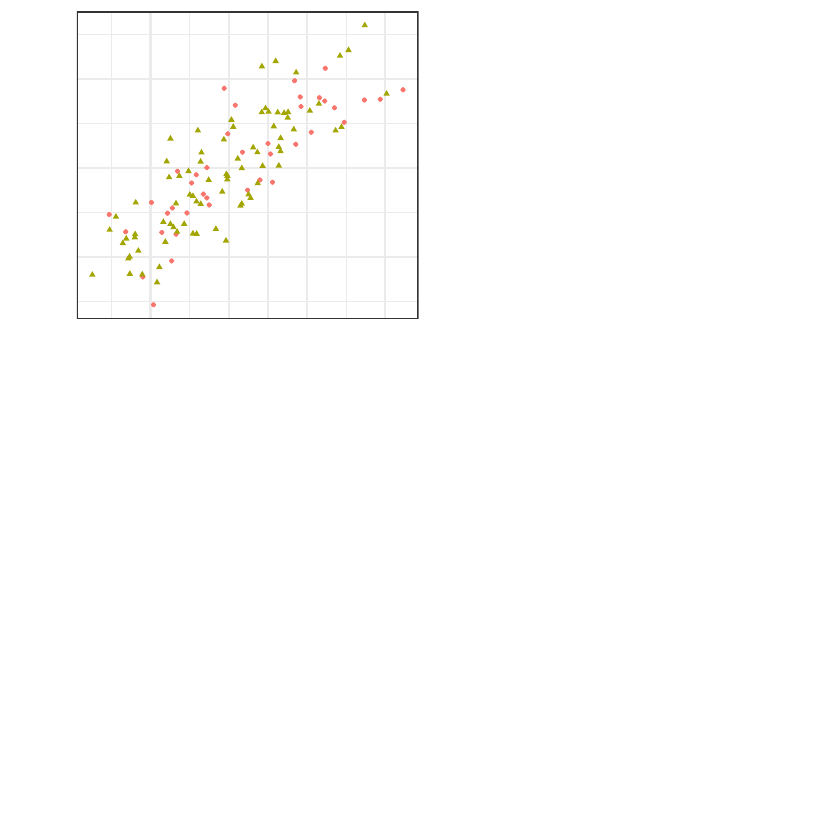
        }
    }
    \caption{Comparison of the simulated data for the linear regression example; true data (red, circle), source when $k=0$ (gold, triangle, top left), source when $k=1$ (green, triangle, top right), source when $k=2$ (blue, triangle, bottom left) and source when $k=3$ (purple, triangle, bottom right).}
    \label{fig:datacomplm}
\end{figure}

\begin{figure}[ph!]
    \centering{
    \resizebox{\textwidth}{!}{
        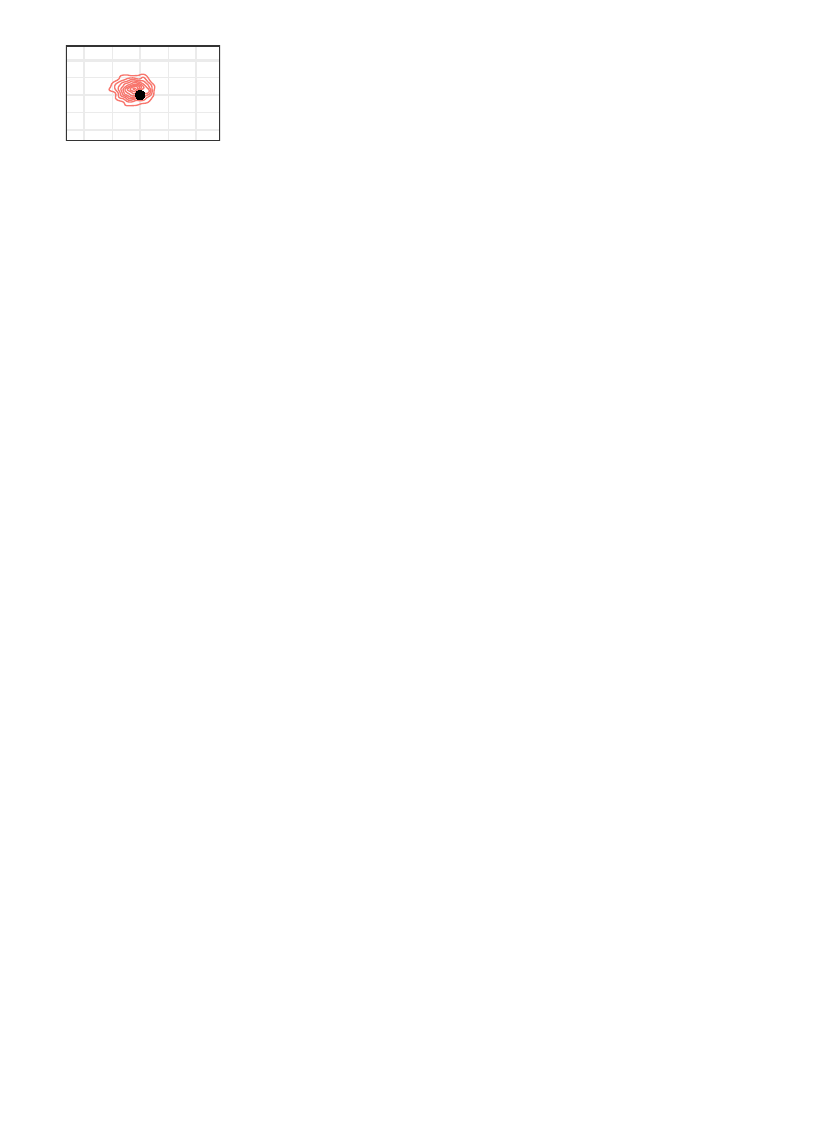
        }
    }
    \caption{Comparison of bivariate ($\beta_0, \beta_1$) posterior density estimate for the linear regression example; true posterior (red), target only posterior (gold), source only posterior (green), Bayesian updating posterior (aqua), FPP posterior (blue) and NPP posterior (purple) on a single dataset.  Each column groups by the value of $k$ and the true value for $\beta_0$ and $\beta_1$ are shown as black dots.}
    \label{fig:posteriorcomplm}
\end{figure}

We let $\Btheta = (\beta_0, \beta_1, \sigma)$, with the target and source parameter values set as follows
\begin{align*}
    \Btheta_\trg &= (5, 3, 2) \\ 
    \Btheta_\src &= (5 + 2\cdot k \cdot \hat{s}, 3 - 2\cdot k \cdot\hat{s}, 2 + 2\cdot k \cdot \hat{s}_{\text{var}}),
\end{align*}
where $k \in \{0, 1, 2, 3\}$ and $\hat{s} \approx 0.15$ is the approximate standard deviation of the true posterior for $\beta_0$ and $\beta_1$ and $\hat{s}_{\text{var}} \approx 0.125$ is the approximate standard deviation of the true posterior for $\sigma$.  This choice of $k$ ensures a balance between allowing the source parameter values to match the target parameter values (where Bayesian updating would be optimal) and lying quite far in the tails of the true posterior based on the target parameter values (where Bayesian updating may perform poorly).  Figure \ref{fig:datacomplm} highlights how it is not immediately obvious when to apply Bayesian transfer learning techniques as the source and target datasets can appear similar.  Further, Figure \ref{fig:posteriorcomplm} shows how the posteriors found for this data differ significantly and how for each level of $k$ a different method performs best and an overview of the results is provided next.


\begin{table}[h]
\centering
\aboverulesep=0ex 
\belowrulesep=0ex
    \caption{Results of the simulation study for the linear regression example: Shown are the average posterior bias, mean squared error (MSE) and 90\% coverage (FCP) for $\overline{\beta}_{0, 1}$ and $\sigma$.  Also shown are the average computed log pointwise predictive density (CLPPD), rank for CLPPD (C-rank), leave-one-out cross-validation (LOO-CV), rank for LOO-CV (L-rank) and chosen $\alpha$ (or posterior median for the NPP) over $100$ independent trials.  Here we compare with the true model for each value for $k$ and highlight the best performing method based on the three ideal metrics in {\color{black}\textbf{black}}.  We highlight the method identified as the best by CLPPD in {\color{blue}\textbf{\textit{blue}}}, and by LOO-CV in {\color{green!65!black}\textbf{\textit{green}}}.}
    \begin{tabular}{*{13}{c}}
         \toprule
         \toprule
         & & \multicolumn{2}{c}{Bias} & \multicolumn{2}{c}{MSE} & \multicolumn{2}{c}FCP & \multirow{2}{*}{CLPPD} & \multirow{2}{*}{C-Rank} & \multirow{2}{*}{LOO-CV} & \multirow{2}{*}{L-Rank} & \multirow{2}{*}{$\alpha$}\\ 
         k & Method & $\overline{\beta}_{0,1}$ & $\sigma$ & $\overline{\beta}_{0,1}$ & $\sigma$ & $\overline{\beta}_{0,1}$ & $\sigma$ & & & & \\
         \hline
         \multirow{6}{*}{0} & True & 0.152 & 0.096 & 0.070 & 0.033 & 0.91 & 0.96 & -84.429 & 3.99  & -85.416 & 3.26  \\ 
         \cmidrule{2-13}
                            & BT & 0.271 & 0.190 & 0.227 & 0.115 & 0.90 & 0.90 & {\color{blue}\textit{\textbf{-83.459}}} & {\color{blue}\textit{\textbf{1.77}}} & -86.518 & 4.64 & 0 \\
                            & BS & 0.178 & 0.117 & 0.100 & 0.046 & 0.89 & 0.92 & -85.643 & 5.80 & -85.671 & 3.74 & -\\
                            & \bf{BU}   & 0.149 & 0.092 & 0.068 & 0.034 & 0.94 & 0.94 & -84.341 & 4.06 & {\color{green!65!black}\textit{\textbf{-85.410}}} & {\color{green!65!black}\textit{\textbf{3.08}}} & 1\\
                            & FPP   & 0.180 & 0.131 & 0.102 & 0.064 & 0.94 & 0.94 & -83.825 & 2.80 & -85.569 & 3.17 & 0.642\\
                            & NPP  & 0.179 & 0.127 & 0.109 & 0.062 & 0.95 & 0.95 & -83.810 & 2.58 & -85.620 & 3.11 & 0.549\\ \hline
         \multirow{6}{*}{1} & True & 0.152 & 0.096 & 0.070 & 0.033 & 0.91 & 0.96 & -84.429 & 3.18  & -85.416 & 2.58  \\ 
         \cmidrule{2-13}
                            & BT & 0.271 & 0.190 & 0.227 & 0.115 & 0.90 & 0.90 & {\color{blue}\textit{\textbf{-83.459}}} & {\color{blue}\textit{\textbf{1.36}}}  & -86.518 & 3.94 & 0 \\
                            & BS & 0.348 & 0.244 & 0.227 & 0.111 & 0.67 & 0.70 & -87.209 & 5.92 & -86.951 & 4.83 & - \\
                            & BU   & 0.227 & 0.194 & 0.113 & 0.071 & 0.79 & 0.70 & -85.303 & 4.53 & -86.011 & 3.38 & 1 \\
                            & FPP   & 0.215 & 0.196 & 0.133 & 0.093 & 0.89 & 0.89 & -84.155 & 2.91 & -85.973 & 3.23 & 0.590 \\
                            & \bf{NPP}  & 0.203 & 0.182 & 0.126 & 0.084 & 0.92 & 0.91 & -84.479 & 3.10 & {\color{green!65!black}\textit{\textbf{-85.885}}} & {\color{green!65!black}\textit{\textbf{3.04}}} & 0.526\\ \hline
         \multirow{6}{*}{2} & True & 0.152 & 0.096 & 0.070 & 0.033 & 0.91 & 0.96 & -84.429 & 2.63  & -85.416 & 2.19  \\ 
         \cmidrule{2-13}
                            & \bf{BT} & 0.271 & 0.190 & 0.227 & 0.115 & 0.90 & 0.90 & {\color{blue}\textit{\textbf{-83.459}}} & {\color{blue}\textit{\textbf{1.16}}} & -86.518 & 3.15 & 0 \\
                            & BS & 0.604 & 0.477 & 0.525 & 0.314 & 0.36 & 0.28 & -89.908 & 5.98 & -89.818 & 5.51 & - \\
                            & BU   & 0.389 & 0.386 & 0.244 & 0.202 & 0.48 & 0.21 & -86.981 & 4.91 & -87.734 & 4.15 & 1 \\
                            & \bf{FPP}   & 0.260 & 0.253 & 0.189 & 0.140 & 0.84 & 0.75 & -84.319 & 2.93 & ${\color{green!65!black}\mathbf{-86.430}}$ & ${\color{green!65!black}\mathbf{2.93}}$ & 0.331 \\
                            & NPP  & 0.271 & 0.279 & 0.192 & 0.154 & 0.84 & 0.67 & -84.924 & 3.39 & -86.714 & 3.07 & 0.416 \\ \hline
         \multirow{6}{*}{3} & True & 0.152 & 0.096 & 0.070 & 0.033 & 0.91 & 0.96 & -84.429 & 2.66  & -85.416 & 1.72  \\ 
         \cmidrule{2-13}
                            & \bf{BT} & 0.271 & 0.190 & 0.227 & 0.115 & 0.90 & 0.90 & {\color{blue}\textit{\textbf{-83.459}}} & {\color{blue}\textit{\textbf{1.17}}}  & {\color{green!65!black}\textit{\textbf{-86.518}}} & {\color{green!65!black}\textit{\textbf{2.58}}}  & 0 \\
                            & BS & 1.025 & 0.754 & 1.236 & 0.674 & 0.05 & 0.05 & -94.893 & 6.00 & -94.665 & 5.93 & - \\
                            & BU   & 0.663 & 0.641 & 0.551 & 0.474 & 0.13 & 0.01 & -90.032 & 4.98 & -90.593 & 4.76 & 1\\
                            & FPP   & 0.271 & 0.287 & 0.218 & 0.173 & 0.88 & 0.80 & -84.320 & 2.86 & -86.708 & 2.88 & 0.129 \\
                            & NPP  & 0.327 & 0.351 & 0.256 & 0.217 & 0.77 & 0.60 & -85.092 & 3.33 & -87.170 & 3.13 &  0.222 \\ \hline
    \end{tabular}
    \label{tab::Ranvar}
\end{table} 

Table \ref{tab::Ranvar} details the results of this example separated by each value of $k$. For clarity, we report on the average of $\beta_0$ and $\beta_1$ which have similar posterior metrics and compare against the true method for each value of $k$.  We see that LOO-CV accurately recommends the appropriate transfer method across all levels of $k$, since the transfer method it prefers aligns well with that chosen by metrics that rely on knowing the true parameter value. In contrast, the CLPPD often recommends the BT method i.e. choosing target only over True in almost all cases.  Below we detail the metrics that show this.

 First, consider the case when $k=0$.  As expected both the True and Bayesian updating methods perform best across all metrics, followed closely by the source only, FPP and NPP methods and the target only method performing the worst.  Next, we consider the case when $k=1$ where FPP and NPP now perform best.  The target only and Bayesian updating methods perform similarly with Bayesian updating having lower MSE but also lower coverage and the source only method already performs noticeably worse.  When $k=2$, the FPP, NPP and target only methods all have similar posterior performance, with the FPP and NPP having slightly lower MSE and coverage --- especially the NPP method.  The Bayesian updating and source only methods both perform worse than the target only method, with higher bias, MSE and significantly lower coverage.  The results for $k=3$ show the target only method slightly outperforming the FPP method, which in turn slightly outperforms the NPP method.  Further, we see that the Bayesian updating method and source only method both have significantly higher bias, MSE and lower coverage.  As is evident from Table \ref{tab::Ranvar}, the best transfer method identified by the LOO-CV approach aligns with the ideal metrics that use a generally unknown true parameter value for all values of $k$.  In contrast, the CLPPD metric erroneously prefers the posterior that uses only the target data for all values of $k$ (even compared to the true posterior).  This demonstrates how our proposed LOO-CV approach can reveal the optimal Bayesian transfer method without access to the true parameter value.
\FloatBarrier

\subsection{Example 2: Weibull Cure Model}\label{subsec52}

As a more realistic example, we use a Weibull cure model as described in \citet{yin2005cure}, with simulated data similar to the melanoma cancer clinical trials E1684 \citep{kirkwood1996interferon} and E1690 \citep{kirkwood2000high}, carried out by the Eastern Cooperative Oncology Group (ECOG).  Both trials consider the effect of Interferon as a treatment for melanoma.  We obtained the data for these two studies from the R package hdbayes \citep{hdbayes}. 

As in \citet{ibrahim2015power}, we model the relapse-free survival time $y_i$ for the~$i$th subject based on the following covariates, a relapse indicator $\nu_i$, a treatment indicator $x_{1, i}$, a sex indicator $x_{2, i}$ and the standardised patient's age in years $x_{3, i}$.  The likelihood function for $n$ observations is
\begin{equation*}
    p(y| \BBeta,\gamma, X) = \prod_{i=1}^n\left(\exp\left(X^T\BBeta\right)f\left(y_i|\gamma\right)\right)^{\nu_i}\exp\left(-\exp\left(X^T\BBeta\right)F\left(y_i|\gamma\right)\right),
\end{equation*}
where $\BBeta = \{\beta_0, \beta_1, \beta_2, \beta_3, \beta_4\}$, with $\beta_0$ an intercept term, $\beta_1, \beta_2$ and $\beta_3$ corresponding to $x_1, x_2$ and $x_3$ respectively, $\beta_4$ is the regression coefficient for the interaction term between $x_2$ and $x_3$ and $f(\cdot)$ and $F(\cdot)$ are the probability density function and cumulative distribution function for the Weibull distribution with parameters $\gamma = \{k, \lambda\}$ given below for a single observation $y$,
\begin{equation*}
    \begin{split}
        f(y|\gamma) &= k \cdot y^{k-1} \cdot \exp\left(\lambda - (y^k)\cdot\exp(\lambda)\right),\\
        F(y|\gamma) &= 1 - \exp\left(-y^k \cdot \exp(\lambda)\right).
    \end{split}
\end{equation*}

The parameter values used to generate data are set to the posterior means found using the E1690 dataset.  For simulating a single observation from the data-generating process, we first simulate the unobserved potentially cancerous cells from a Poisson distribution,~$C\sim\text{Poi}\left(\exp\left(X\BBeta\right)\right)$.  The mean parameter for the Poisson depends on the design matrix $X$, which we simulate based on the summaries of the E1690 dataset.  That is, we generate $x_1 \sim \text{Ber}\left(0.511\right)$, $x_2 \sim \text{Ber}\left(0.397\right)$ and $x_{3, i} = h(s_i),\ s \sim \mathcal{N}\left(0, 0.6^2\right)$, where $h(s)$ is a location-scale transformation so that $x_3$ is standardised to a $\mathcal{N}(0, 1)$ distribution.  Then we simulate a single relapse time $y$ to be the minimum of $C$ realisations from a Weibull distribution,
\begin{equation*}
    \begin{split}
        y &= \min\left(\{z_j\}_{j=1}^{C}\right) \\ 
        z_j &\sim \text{ Wei}(k, \lambda), \quad \mbox{ for } j = 1,\ldots,C.
    \end{split}
\end{equation*}
We set $\nu =0$ when $C=0$ or the simulated relapse time is greater than the right censor value of $5.5$, otherwise, we consider a relapse to have occurred and set $\nu = 1$.  As in the previous example, we draw $40$ data points for the target data; however, to more realistically simulate a related study we generate $300$ data points for the source data.  Both sets of data are generated with the following parameter values,
\begin{equation*}
    \begin{split}
        \Btheta_\trg &= \left(0.163, -0.299, 0.120, -0.287, 0.276, 1.103, -0.538 \right)^T \\
        \Btheta_\src &= \Btheta_\trg + 2 \cdot k \cdot \hat{s}, 
    \end{split}
\end{equation*}
where $k \in \{0, 1, 2, 3\}$, $\hat{s} = (0.115, 0.160, 0.066, 0.190, 0.270, 0.064, 0.104)$ is the vector of estimated standard deviation values for each marginal posterior based on the target data, $\Btheta = \{\beta_0, \beta_1, \beta_2, \beta_3, \beta_4, k, \lambda\}$ with $\Btheta_\trg$ and $\Btheta_\src$ indicating the target and source parameter values, respectively.  The results are discussed next. 

\begin{table}[h]
\centering
\aboverulesep=0ex 
\belowrulesep=0ex
    \caption{Results of the simulation study for the Weibull cure model: Shown are the average posterior bias, mean squared error (MSE) and 90\% coverage (FCP) for the average of  $\Btheta = \{\beta_0, \beta_1, \beta_2, \beta_3, \beta_4, k, \lambda\}$.  Also shown are the average computed log pointwise predictive density (CLPPD), rank for CLPPD (C-rank), leave-one-out cross-validation (LOO-CV), rank for LOO-CV (L-rank) and chosen $\alpha$ (or posterior median) over $100$ independent trials.  Here we compare with the true model for each value for $k$ and highlight the best performing method based on the three ideal metrics in {\color{black}\textbf{black}}.  We highlight the method identified as the best by CLPPD in {\color{blue}\textbf{\textit{blue}}}, and by LOO-CV in {\color{green!65!black}\textbf{\textit{green}}}.}

    \begin{tabular}{*{10}{c}}
         \toprule
         \toprule
         k & Method & Bias & MSE &  FCP & CLPPD & C-Rank & LOO-CV & L-Rank & $\alpha$\\
         \hline
         \multirow{6}{*}{0} & True & 0.125 & 0.056 & 0.90 & -53.266 & 3.9 & -53.719 & 3.0 \\ \cmidrule{2-10}
                            & BT & 0.486 & 1.115  & 0.91 & {\color{blue}\textit{\textbf{-50.728}}} & {\color{blue}\textit{\textbf{1.1}}} & -57.166 & 5.7 & 0 \\
                            & \bf{BS} & 0.131 & 0.067 & 0.90 & -54.129 & 5.763 & -53.675 & 2.979 & - \\
                            & \bf{BU}   & 0.128 & 0.059 & 0.90 & -53.335 & 4.320 & {\color{green!65!black}\textit{\textbf{-53.652}}} & {\color{green!65!black}\textit{\textbf{2.773}}} & 1 \\
                            & FPP   & 0.141 & 0.100 & 0.93 & -52.669 & 3.320 & -53.752 & 3.134 & 0.747\\
                            & NPP  & 0.144 & 0.116 & 0.97 & -52.557 & 2.557 & -53.836 & 3.402 & 0.571\\ \hline
         \multirow{6}{*}{1} & True & 0.125 & 0.056 & 0.90 & -53.266 & 3.7 & -53.719 & 1.5 \\ \cmidrule{2-10}
                            & BT & 0.486 & 1.115 & 0.91 & {\color{blue}\textit{\textbf{-50.728}}} & {\color{blue}\textit{\textbf{1.5}}} & -57.166 & 3.6 & 0 \\
                            & BS   & 0.271 & 0.144 & 0.45 & -67.716 & 6.000 & -66.956 & 5.866 & - \\
                            & BU   & 0.232 & 0.108 & 0.51 & -61.920 & 5.000 & -62.454 & 4.629 & 1 \\
                            & \bf{FPP}   & 0.307 & 0.512 & 0.94 & -51.050 & 2.021 & {\color{green!65!black}\textit{\textbf{-55.863}}} & {\color{green!65!black}\textit{\textbf{2.577}}} & 0.048\\
                            & \bf{NPP}  & 0.287 & 0.461 & 0.91 & -51.943 & 2.845 & -56.354 & 2.814 & 0.095\\ \hline
         \multirow{6}{*}{2} & True & 0.125 & 0.056 & 0.90 & -53.266 & 3.9, & -53.719 & 1.4 \\ \cmidrule{2-10}
                            & BT & 0.486 & 1.115 & 0.91 & {\color{blue}\textit{\textbf{-50.728}}} & {\color{blue}\textit{\textbf{1.8}}} & -57.166 & 3.4 & 0 \\
                            & BS & 0.538 & 0.430 & 0.07 & -140.08 & 6.000 & -134.93 & 6.000 & - \\
                            & BU   & 0.368 & 0.219 & 0.30 & -83.315 & 5.000 & -84.075 & 4.979 & 1 \\
                            & \bf{FPP}   & 0.402 & 0.833 & 0.93 & -50.789 & 1.979 & {\color{green!65!black}\textit{\textbf{-56.352}}} & {\color{green!65!black}\textit{\textbf{2.588}}} & 0.013\\
                            & \bf{NPP}  & 0.391 & 0.804 & 0.92 & -50.904 & 2.320 & -56.371 & 2.680 & 0.017\\ \hline
         \multirow{6}{*}{3} & True & 0.125 & 0.056 & 0.90 & -53.266 & 3.9 & -53.719 & 1.4 \\ \cmidrule{2-10}
                            & BT & 0.486 & 1.115 & 0.91 & {\color{blue}\textit{\textbf{-50.728}}} & {\color{blue}\textit{\textbf{1.6}}} & -57.166 & 3.3 & 0 \\
                            & BS & 0.806 & 0.901 & 0.02 & -359.59 & 6.000 & -342.67 & 6.000 & - \\
                            & BU   & 0.437 & 0.314 & 0.34 & -102.24 & 5.000 & -103.49 & 5.000 & 1 \\
                            & \bf{FPP}   & 0.439 & 0.972 & 0.92 & -50.844 & 2.000 & -56.636 & 2.691 & 0.007\\
                            & \bf{NPP}  & 0.430 & 0.936 & 0.91 & -50.927 & 2.256 & {\color{green!65!black}\textit{\textbf{-56.566}}} & {\color{green!65!black}\textit{\textbf{2.639}}} & 0.009\\ \hline
    \end{tabular}
    \label{tab::Weib}
\end{table}

From Table \ref{tab::Weib} it is clear that across all levels of $k$, LOO-CV accurately identifies the best transfer method as it aligns with the method selected using the metrics that exploit the true target parameter value.  The drawback of CLPPD is again highlighted under the Bayesian transfer learning setting, as it consistently chooses the BT method over the true method.  This can be clearly seen when $k=0$, under this setting the BT method performs significantly worse across all three of the ideal metrics and still CLPPD identifies it as the best method.

As in the previous example, when $k=0$ the Bayesian updating method performs best, followed closely by the source only, FPP and NPP methods.  Compared to the linear regression, the data are less informative since the data represent an order statistic of the Weibull distribution, and can be censored.  Furthermore, the source dataset is larger than the previous example.  Therefore, the target only method performs poorly in this simulation study, with a significantly higher MSE and bias compared to the previous example.  When $k=1$ in this setting, the FPP and NPP approaches perform best, having a bias and MSE similar to the Bayesian updating method, but much better coverage.  The source only and Bayesian updating methods have significantly worse coverage than the target only method.  The results for $k=2$ and $k=3$ show that the FPP and NPP methods slightly outperform the target only method.  Again, the Bayesian updating and source only methods perform poorly, especially for coverage where the source only method has almost $0\%$ coverage. 

\section{Discussion}\label{sec6}

Bayesian transfer learning methods incorporate related source data to improve inference on the target.  Previous methods offer no means of identifying when Bayesian transfer learning methods should be used over completely discarding or incorporating the source data.  Additionally, previous power prior methods offer no computationally efficient way to evaluate both the FPP and NPP posteriors.  In this work, we have proposed using posterior predictive checks to address the model selection problem.  Further, we compared the performance of posterior predictive checks, namely the CLPPD and LOO-CV, for choosing the appropriate transfer method.  We have presented a computationally efficient framework, TSMC, to implement multiple power prior approaches.  TSMC uses two adaptive SMC schedules to sample the relevant sequence of posteriors and approximate the corresponding normalising constants.  Finally,  we show how the TSMC framework enables us to perform LOO-CV on the doubly intractable NPP easily. 

Based on our simulation studies it is clear that posterior predictive checks are an accurate way to evaluate the usefulness of transfer learning techniques.  That is, LOO-CV accurately identifies the best performing transfer method.  However, care must be taken with the choice of posterior predictive checks as evidenced by the poor performance of computing the CLPPD on the target data.  It appears that the CLPPD is biased by fitting and evaluating on the target data, and thus prefers methods more influenced by the target data.  The CLPPD even preferred the target only posterior over the true posterior, which is based on the true parameter value with a larger sample size.  One limitation of LOO-CV is the increased computational cost of evaluating $n$ posterior distributions, associated with leaving one of the target observations out for each element in the target dataset.  Fortunately, with the TSMC framework, we can easily apply importance sampling to efficiently obtain these posteriors --- including those required in the doubly intractable NPP.

The simulation studies revealed that when $\trg$ and $\src$ are moderately related the power prior outperforms the target only and Bayesian updating methods.  Although the TSMC framework adaptively chooses the appropriate amount of transfer it is still beneficial to run all four methods (BT, BU, FPP and NPP) and utilise the LOO-CV posterior predictive check to identify the best-performing method.  In our examples, we have found that the additional computational cost of evaluating LOO-CV within the TSMC framework is reasonable.  


Under the Bayesian transfer learning setting, it might be desirable to update from more than one source dataset.  Our current framework does not consider this setting.  However, \citet{gravestock2019power} propose to incorporate multiple source datasets by treating them as independent and estimating a unique transfer parameter for each. Therefore, an idea for future exploration could be to incorporate each source dataset individually by applying our TSMC framework sequentially (pairwise).  We could start by applying TSMC to the first source dataset and target dataset.  The resulting posterior can then be used as the target posterior, fixing $\alpha$, for the next source dataset.  We could continue this pairwise application of TSMC until we finally incorporate all the source datasets.  This scheme could allow useful information to be transferred from multiple source datasets.

Two alternative Bayesian transfer learning approaches to the power prior are the commensurate prior, which makes use of a spike and slab prior, and MAPA, which utilises a robust mixture distribution.  Unfortunately, for these methods, the choice of proposal distribution, which allows efficient posterior sampling, is not clear \citep{biswas2022scalable}.  Therefore, future work could consider effective proposal distributions or a more computationally efficient framework to evaluate the commensurate prior and MAPA methods.  Fortunately, the LOO-CV metric we presented is still applicable to assess the performance of these methods.

One key limitation of the NPP, which we do not attempt to address in this paper, is the choice of prior for $\alpha$ \citep{pawel2022normalized}.  A standard Bayesian approach is to use an uninformative prior.  Such an uninformative prior neglects to account for the increased influence the additional source data provides.  Consider our second simulation study where there are $40$ target and $300$ source data points. Under this setting, the likelihood evaluation of the source data will dominate that of the target data.  Further, this influence will only compound as we incorporate multiple source datasets.  Future research could use the LOO-CV metric to evaluate different prior choices for the NPP.

\backmatter


\bmhead{Acknowledgements}

AB and CD were supported by an Australian Research Council Future Fellowship (FT210100260). \\
DJW is supported by an Australian Research Council Early Career Researcher Award (DE250100396). \\
JJB was supported by the European Union under the (2023-2030) ERC Synergy grant 101071601 (OCEAN). Views and opinions expressed are however those of the author only and do not necessarily reflect those of the European Union or the European Research Council Executive Agency. Neither the European Union nor the granting authority can be held responsible for them. \\
The authors would like to thank the two anonymous referees for their helpful comments, which led to improvements in this paper.




\noindent

\begin{appendices}

\end{appendices}


\bibliography{bibliography}

\end{document}